\documentstyle[11pt,amssymb,amsfonts,amsthm,amsmath,amssymb,amscd,amstext,epsfig]{article}

\def\ben{\begin{equation}}
\def\een{\end{equation}}
\def\half{{\textstyle{1\over2}}}

\let\a=\alpha \let\b=\beta   
   
\let\l=\lambda

\let\pa=\partial
\def\be{\begin{equation}}
\def\ee{\end{equation}}
\def\ba{\begin{array}}
\def\ea{\end{array}}

\def\dalemb#1#2{{\vbox{\hrule height .#2pt
        \hbox{\vrule width.#2pt height#1pt \kern#1pt
                \vrule width.#2pt}
        \hrule height.#2pt}}}

\newcommand{\bea}{\begin{eqnarray}}
\newcommand{\eea}{\end{eqnarray}}
\newcommand{\tr}{{\rm tr} }

\def\R{{{\Bbb R}}}

\def\calD{{\mathcal{D}}}

\begin{document}

\begin{flushright}
NSF-KITP-08-68 \\
WIS/10/08-MAY-DPP \\
arXiv:0805.4658 [hep-th]
\end{flushright}

\begin{center}
\vspace{1cm} { \LARGE {\bf Multi-matrix models and emergent geometry}}

\vspace{1.1cm}

David E. Berenstein$^\sharp$, Masanori Hanada$^\natural$ and Sean
A. Hartnoll$^\flat$

\vspace{0.8cm}

{\it $^\sharp$ Department of Physics, University of California\\
     Santa Barbara, CA 93106-9530, USA }

\vspace{0.5cm}

{\it $^\natural$ Department of Particle Physics, Weizmann Institute of Science\\
     Rehovot 76100, Israel }

\vspace{0.5cm}

{\it $^\flat$ KITP, University of California\\
     Santa Barbara, CA 93106-4030, USA }

\vspace{0.8cm}

{\tt dberens@physics.ucsb.edu, masanori.hanada@weizmann.ac.il, hartnoll@kitp.ucsb.edu} \\

\vspace{1.2cm}

\end{center}

\begin{abstract}
\noindent
Encouraged by the AdS/CFT correspondence, we study emergent local
geometry in large $N$ multi-matrix models from the perspective of
a strong coupling expansion. By considering various solvable
interacting models we show how the emergence or non-emergence of
local geometry at strong coupling is captured by observables that
effectively measure the mass of off-diagonal excitations about a
semiclassical eigenvalue background. We find emergent geometry at
strong coupling in models where a mass term regulates an infrared
divergence. We also show that our notion of emergent geometry can be usefully
applied to fuzzy spheres. Although most of our results are analytic, we have
found numerical input valuable in guiding and checking our
results.
\end{abstract}

\pagebreak
\setcounter{page}{1}

\section{Introduction}

The emergence of geometry has long been recognised as a key issue
in quantum gravity. The more recent discovery of the AdS/CFT
correspondence \cite{Maldacena:1997re} has indicated a
complementary viewpoint: Certain large $N$ field theories are most
simply described starting from a higher dimensional dynamical
geometry. The geometry is not apparent in the weak coupling
(Lagrangian) description of the quantum field theory, and in this
sense is emergent. The emergence of spacetime can be thought of as
a precise realisation of 't Hooft's insight \cite{'t Hooft:1973jz}
that large $N$ gauge theories are string theories, together with
the fact that string theory describes a theory of quantum
geometry.

The best understood cases of emergent geometry from field theory
begin with D branes in a pre-existing
geometry\footnote{Early pre-stringy observations
of emergent geometry include for instance \cite{Bhanot:1982sh}.}.
The geometry `dual' to the strongly coupled field theory
on the D brane worldvolume is then obtained by computing the
gravitational backreaction of the D branes
\cite{Maldacena:1997re, Itzhaki:1998dd}. These theories are
often supersymmetric and, in the best understood cases, conformal.
One would want a much more comprehensive understanding of how
geometry emerges, and the relevant situations where it can be
applied. In essence, one needs to find a way to solve a key aspect
of the field theory dynamics and find geometry.

There is ample reason to believe that the large $N$ expansion
might be relevant for QCD, the theory of the strong interactions,
and one would like to have a first principles approach to
calculating the corresponding dual geometry (or dual string
theory). If this is understood, exploiting the geometrical
information may allow us to describe the strong coupling dynamics
of the theory in a more economical way. The strong interactions
are not supersymmetric and, in the regime of strong coupling, they
are not conformal either.

In this paper we would like to study emergent geometry from first
principles in a simplified setting. We will study large $N$
systems in zero dimensions, with and without supersymmetry. We
will solve the multi-matrix models exactly in certain limits, and
look for emergent geometry.

The simplest emergence of geometry from large $N$ matrices occurs
in a Gaussian matrix model for a single matrix. In the large $N$
limit, the integral is dominated by a saddle point in which the
eigenvalues of the matrix are distributed in a semicircle
\cite{wigner}. This semicircle can be thought of as a continuum
geometry that emerges at large $N$. This very simple example
already illustrates an important theme for us. The emergent
geometry is possible because the Gaussian mass term balances the
repulsive inter-eigenvalue force. The Gaussian model may be
generalised by introducing interaction terms for the matrix, and
an elegant mathematical theory allows us to find the eigenvalue
distribution in that case \cite{Brezin:1977sv}.

A tractable step beyond single matrix models are normal matrix
models, in which a Hermitian matrix and its Hermitian conjugate commute
with one another. These models show an emergent geometry
at large $N$ which describes a two dimensional droplet \cite{Chau:1991gj}.
Again there is an elegant mathematical framework to describe these
models, see for instance \cite{Teodorescu:2004qm, Chau:1997pr}.

For general multi-matrix models, even with just two matrices, the
question of whether there is or not an emergent geometry is
difficult and a developed framework is lacking. Unlike the case of
single matrix or normal matrix models, these systems can rarely
\cite{Itzykson:1979fi} be solved exactly even at large $N$. Some
recent numerical work can be found in \cite{Krauth:1998xh,
Krauth:1998yu, Krauth:1999qw, Krauth:1999rc, Hotta:1998en,
Ambjorn:2000bf}. However, many of the cases of most interest are
numerically problematic because of the sign problem in certain
supersymmetric theories.

The AdS/CFT correspondence \cite{Maldacena:1997re} suggests that a
natural organising principle is a strong coupling limit. For the
correspondence to work, the degrees of freedom in the field theory
need to `geometrise' at strong coupling in order to reproduce the
dual higher dimensional (dynamical) spacetime. There are various
discussions of emergent geometry in AdS/CFT in the literature,
including \cite{Dorey:1999pd, Rey:2005cn, Chen:2006ps}. However,
only recently has an attempt been made to systematise the
emergence of geometry as a consequence of the strong coupling
limit \cite{Berenstein:2005aa}.

With multiple matrices, it seems that a key aspect of an emergent
classical geometry is that the matrices commute with each other in
the large $N$ limit. In this way the typical $N^2$ degrees of
freedom of matrices get effectively reduced to order $N$ degrees
of freedom at low energies. The collective description of these
low energy degrees of freedom can often be given in terms of a
joint eigenvalue distribution for several matrices. It is the
geometrical description of this eigenvalue distribution that
produces the emergent geometry. One objective of this paper is to
make the notion of commuting matrices more precise. In
\cite{Berenstein:2005aa} it was proposed that the zero modes of the
six scalar fields of ${\mathcal{N}}=4$ super Yang-Mills theory on
a spatial $S^3$ commuted at strong coupling. From this proposal
one can show that the joint eigenvalue distribution of these six
matrices forms an $S^5$ that should be identified with the
geometric $S^5$ that arises in the $AdS_5 \times S^5$ of the dual
IIB string theory.

The proposal of \cite{Berenstein:2005aa} was subsequently
generalised to describe orbifolds of the ${\mathcal{N}}=4$ theory
\cite{Berenstein:2005ek} and also to ${\mathcal{N}}=1$
theories \cite{Berenstein:2007wi, Berenstein:2007kq}. Various
successful checks of the proposal were performed in
\cite{Berenstein:2005jq, Berenstein:2007kq, Berenstein:2008jn}.
However, given that the theory involves infinitely many coupled
matrices with comparable masses, it is difficult to prove the
validity of the truncation to just six fields. Furthermore, the
theories considered have all been superconformal and somewhat
similar to the maximally symmetric Yang-Mills theory in four
dimensions.

In this paper we will look at models in which we can show directly
and unambiguously whether the matrices commute or not at strong
coupling. The paper is organised as follows. We will solve a
succession of multi-matrix models in the strong coupling limit. At
each step, we shall check our analytic results with numerics.
Although many of our final results are analytic, the interplay
with numerics has been a very important guide towards these
results. In a later section we consider some models that we
cannot solve analytically. Finally, we also show how our notion
of emergent geometry can be applied to fuzzy spheres.
The concluding discussion includes a
summary of our results. The main point we emphasise is that higher
dimensional emergent geometry can arise naturally in models where
a mass term regularises an infrared divergence of a massles model.
In these cases, at strong coupling, the eigenvalues are
sufficiently spread out that off-diagonal modes do not contribute
to low energy dynamics.

\section{A two matrix model at strong coupling}
\label{sec:parabolic}

\subsection{Solving the model: Parabolic distribution}

We begin by considering a two matrix model that can and has
\cite{Kazakov:1998ji} been solved exactly at all couplings.
We could read off most of the quantities we need from
\cite{Kazakov:1998ji}. Instead we will solve the model in a more
low-tech way, that is geared towards the more general issues we
would like to understand at strong 't Hooft coupling. We will see
that this model can be re-interpreted as an emergent two
dimensional geometry at strong coupling.

Consider the two Hermitian matrix model
\be\label{eq:original}
Z = \int \calD X \calD Y e^{- \tr X^2 -
\tr Y^2 + g^2 \tr [X,Y]^2}\,.
\ee
Because this integral is quadratic in both $X$ and $Y$, we can
diagonalise $X$ and then perform the $Y$ integral exactly. The
partition function becomes
\bea\label{eq:z2}
Z & = & \int dx_1 \ldots dx_N e^{-
\sum_i x_i^2} \prod_{i \neq j} \frac{x_i -
x_j}{\sqrt{1 + g^2 (x_i - x_j)^2}}  \,, \\
& = & \int dx_1 \ldots dx_N e^{-
\sum_i x_i^2 + \frac{1}{2} \sum_{i \neq j} \log(x_i-x_j)^2 - \frac{1}{2} \sum_{i \neq j} \log[1 +
g^2 (x_i-x_j)^2]}\,.
\eea
The first logarithm appearing here is the standard Vandermonde
determinant arising upon diagonalising $X$. We can look for a
large $N$ saddle point to this integral. The saddle point
equations of motion are
\be\label{eq:saddle}
x_i = \sum_{j \neq i} \frac{1}{(x_i - x_j) [1 + g^2 (x_i -
x_j)^2]}\,.
\ee
These equations have been solved at large $N$ by
Kazakov-Kostov-Nekrasov \cite{Kazakov:1998ji}, somewhat
implicitly. Here we shall be interested in the strong 't Hooft
coupling limit $\lambda = g^2 N
\gg 1$, and shall take a less sophisticated approach.

In the large $N$ limit, the saddle point equation becomes
\be\label{eq:integral}
x = \int \frac{\rho(y) dy}{(x-y) [1 + g^2 (x-y)^2]}
\,.
\ee
We took the continuum limit $\sum \to \int \rho(x) dx$ and
introduced the eigenvalue density $\rho(x)$. We are working with
the normalisation $\int \rho(x) dx = N$. As usual, a principal
value is understood in the integral (\ref{eq:integral}).

\begin{figure}[h]
\begin{center}
\epsfig{file=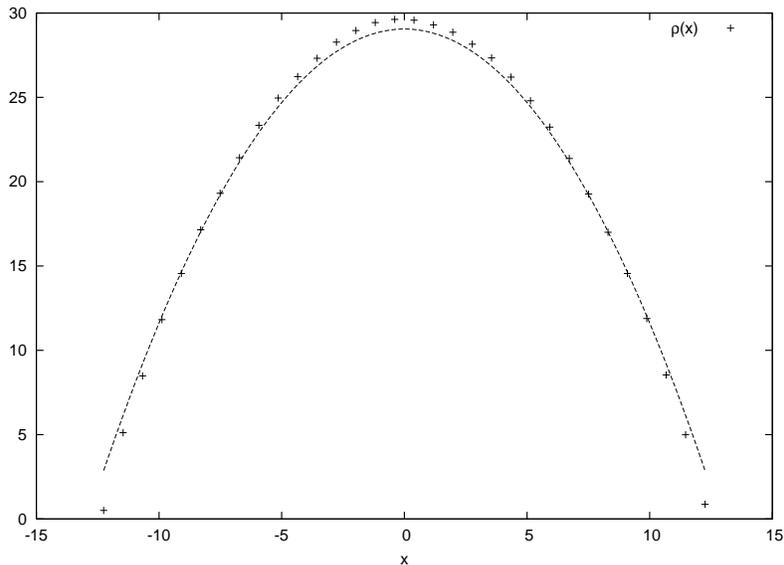,width=3in,angle=270,trim=0 0 0 0}%
\end{center}
\caption{Numerically simulated eigenvalue distribution solution to
(\ref{eq:saddle}) with $N=500$ and $\lambda = 600$,
together with the theoretical result (\ref{eq:solution}), which has
$L=12.91$ in this case.
  \label{fig:parabola}}
\end{figure}

We can solve this equation at strong 't Hooft coupling by first
noting the general result
\be\label{eq:general}
{\mathcal{P}} \int_{-1}^1 \frac{\rho(y) dy}{(x-y) [1+ J^2
(x-y)^2]} = - \frac{\pi \rho'(x)}{J}+ \cdots \quad
\text{as} \quad J \to\infty \,.
\ee
One can check that indeed
\be\label{eq:expand}
{\mathcal{P}} \int_{-1}^1 \frac{(1-y^2) dy}{(x-y) [1+ J^2
(x-y)^2]} = \frac{2 \pi x}{J}+ \cdots \quad \text{as} \quad J
\to\infty
\,.
\ee
The key point is that this result is linear in $x$. From this
integral, it follows that in the strong 't Hooft coupling limit,
$\lambda = g^2 N \gg 1$, the solution to the integral equation
(\ref{eq:integral}) is the parabolic distribution
\be\label{eq:solution}
\rho(x) = \frac{3 N  (L^2 - x^2)}{4 L^3} \,,
\ee
with
\be\label{eq:L}
L = N^{1/2} \frac{(3\pi)^{1/3}}{2^{1/3}}
\frac{1}{\lambda^{1/6}} \,.
\ee
The correctness of this solution can be verified by plugging it
into the equation (\ref{eq:integral}), performing the integral and
then taking the large $\lambda$ limit. We have also simulated
the eigenvalue partition function (\ref{eq:z2}) numerically
using a Hybrid Monte-Carlo algorithm. The resulting distribution is
shown for $N=500$ and $\lambda = 600$ in figure \ref{fig:parabola}.
We stored 10000
configurations and determined the distribution $\rho(x)$
from the distribution of $N \times 10000$ points. The plot shows
that the eigenvalue distribution is indeed close to our
theoretical result (\ref{eq:solution}). An
analogous plot at lower $\lambda$ (say $\lambda=20$) shows
clear deviations away from being a parabola. Note that the $N$ appearing
in these computations is just a discretisation of the integral equation
(\ref{eq:integral}). Therefore, for these purposes, we can take $\lambda$ to be larger than
$N$ if we wish, without upsetting the 't Hooft limit.

We should note however, that the solution (\ref{eq:solution}) is
not correct very near to the endpoints of the distribution, $x
\sim \pm L$. More specifically, it is not correct for $(L^2 - x^2)
\lesssim L^2/\lambda^{1/3}$. This occurs because there are higher
order corrections to (\ref{eq:expand}) of the form
$1/[(1-x^2)J]^n$, with $n>0$, which become large near the
endpoints. However, this represents a vanishingly small fraction
of the support of the eigenvalue density and so should not be
important for many observables. Specifically
\be
\frac{\text{\# incorrect}}{\text{\# total}} \sim
\frac{1}{\lambda^{2/3}} \to 0\,.
\ee
Furthermore, by going to very large $\lambda$ we have checked that
the result for the width $L$ in (\ref{eq:L}) agrees excellently
with numerics, including the prefactor $(3\pi/2)^{1/3} \approx
1.68$.

For a typical pair of eigenvalues, the solution
(\ref{eq:solution}) implies that
\be
g^2 (x-y)^2 \sim \lambda L^2/N \sim \lambda^{2/3} \gg 1 \,.
\ee
This observation suggests that the matrices $X$ and $Y$ are
effectively commuting at strong coupling, because most of the off diagonal
modes are parametrically more massive than the diagonal modes.
Recall that the off diagonal mode connecting the $i$th and $j$th
eigenvalue has mass $1+ g^2 (x_i - x_j)^2$, as we used for
instance in evaluating the determinant (\ref{eq:z2}). We shall
make this notion of commutativity more precise in the following
subsection.

It is now easy to evaluate, for instance,
\be\label{eq:x2}
\frac{\tr X^2}{N} = \int_{-L}^L \frac{3 x^2 (L^2-x^2))}{4 L^3} dx =
\frac{L^2}{5} = \frac{(3\pi)^{2/3} N}{5 \cdot 2^{2/3} \lambda^{1/3}} + \cdots \,.
\ee
We can note that this agrees exactly with the strong coupling
result by Kazakov-Kostov-Nekrasov, see \cite{Kazakov:1998ji}
equation (6.29).

\subsection{Commutators and criteria for locality}
\label{sec:2matrix}

Let us now consider observables that depend on both matrices $X$
and $Y$ in the two matrix model. Our objective here is to show
more quantitatively that the two matrices commute at strong
coupling. Our comment in the previous subsection about the large mass
of off-diagonal modes is insufficient in itself. In the framework we are
using so far, in which $X$ is diagonalised and $Y$ treated exactly,
the off-diagonal modes do in fact make significant contributions to generic
observables involving $Y$. We shall see this shortly.
When $X$ and $Y$ are treated on an unequal footing, it is best to
discuss basis-independent quantities.
In particular, we are interested in the following combinations
\be
\tr (XYXY) \,, \qquad \tr (X^2 Y^2) \,, \qquad \tr [X,Y]^2 = 2 \left( \tr
(XYXY) - \tr (X^2 Y^2) \right) \,.
\ee
Obviously the commutator square should have information on how
close to commuting are two sets of matrices. The other two
combinations have different large $N$ behavior in free matrix
models. The term $\tr(XYXY)$ would vanish at the planar level, and
$\tr(X^2Y^2)$ would not. We would consider the case of completely
uncorrelated matrices to mean that the matrices are non-commuting.
The ratio
\begin{equation}
r= \frac{\tr(XYXY)}{\tr(X^2Y^2)} \,,
\end{equation}
would then serve as an order parameter that tells us something
about the correlation of the matrices. To leading order in $1/N$
it vanishes in the Gaussian model.

It is easy to show that $|r|\leq1$ for Hermitian matrices. It
results from a simple manipulation of the following two
inequalities:
\begin{eqnarray}
\tr([X,Y]^2) \leq 0 \,, \\
\tr(\{X,Y\}^2)\geq 0 \,.
\end{eqnarray}
These follow from the fact that $[X,Y]$ and $\{X,Y\}$ are
anti-Hermitian and Hermitian respectively. For matrices that
commute, we find that $r=1$, while for matrices that anticommute
we would find that $r=-1$. In the case $r=-1$, the square of the
matrices would commute with each other, so there is also a lot of
order in the eigenspaces of the matrices.

When we consider our two matrix model, the easiest of these to compute is the commutator squared.
Specifically
\be
\tr [X,Y]^2 = \frac{N}{Z} \frac{\pa Z}{\pa \lambda} = N \frac{\pa \log Z}{\pa \lambda} \,.
\ee
Here $\log Z$ should be evaluated on the large $N$ saddle
\be
\log Z = - \int \rho(x) x^2 dx + \int dx dy \rho(x) \rho(y)
\left( \frac{1}{2} \log(x-y)^2 - \frac{1}{2} \log [1+ g^2 (x-y)^2] \right)\,.
\ee
Using our parabolic solution (\ref{eq:solution}) we obtain to
leading order at large $\lambda$
\be\label{eq:commutator}
\tr [X,Y]^2 = - \frac{N^3}{2 \lambda} + \cdots \,.
\ee
This is the correct answer for the commutator to leading order in
large $\lambda$. It is clear from the computation that higher
order corrections to the eigenvalue distribution give subleading
contributions.

To get the other traces we can introduce a source $J_{ij}$ into
the action for the $Y_{ij}$ component of $Y$. Let
\be
Z[J] = \int \calD X \calD Y e^{- \tr X^2 -
\tr Y^2 + g^2 \tr [X,Y]^2 +  \tr J Y} \,.
\ee
Diagonalising the $X$ matrix, this becomes
\be
Z[J] = e^{\frac{1}{4} \sum_{i \neq j} |J_{ij}|^2 [1+ g^2 (x_i -
x_j)^2]^{-1} } Z[0] \,.
\ee
Thus
\be\label{eq:XYXY}
\tr (XYXY) = \sum_{i,j} \frac{x_i x_j}{Z[0]} \left. \frac{\delta^2 Z[J]}{\delta
J_{ij} \delta J_{ji}} \right|_{J=0}= \int \frac{dx dy \rho(x)
\rho(y) x y}{2 [1 + g^2 (x-y)^2]} \,,
\ee
and similarly
\be\label{eq:XXYY}
\tr (X^2 Y^2) = \int \frac{dx dy \rho(x) \rho(y) x^2}{2 [1 + g^2
(x-y)^2]} \,.
\ee
It is easy to evaluate these integrals using our large $N$
distribution (\ref{eq:solution}) to obtain at large $\lambda$
\bea\label{eq:fourpoint}
\tr (XYXY) = \frac{3 \pi}{70} \left(\frac{3 \pi}{2} \right)^{1/3}
\frac{N^3}{\lambda^{2/3}} - \frac{3 N^3}{8 \lambda} + \frac{\a N^3}{\lambda} + \cdots \,,
\\
\tr (X^2 Y^2) = \frac{3 \pi}{70} \left(\frac{3 \pi}{2} \right)^{1/3}
\frac{N^3}{\lambda^{2/3}} - \frac{N^3}{8 \lambda} + \frac{\a N^3}{\lambda} + \cdots \,.
\eea
In these expressions we have included an unknown contribution $\a
N^3/\lambda$ that comes from the leading correction to the
parabolic eigenvalue distribution (\ref{eq:solution}) in the large
$\lambda$ limit. We know that the contribution has to be equal in
the two expressions because taking their difference recovers our
previous result for the commutator (\ref{eq:commutator}). Recall
that our expression for the commutator did not depend on
corrections to the eigenvalue distribution. The ratios we are
about to consider do not depend on $\a$. For completeness, in
Appendix A we show that
\be
\a = - \frac{1}{40} \,.
\ee
We checked the leading order results for the commutators
numerically, simulating the full partition function
(\ref{eq:original}).

We can now ask how these observables capture the commutativity of
the matrices at strong coupling. The most na\"ive object to look
at would be
\be\label{eq:one}
\frac{N \tr [X,Y]^2}{\tr X^2 \tr Y^2} = - \frac{25}{2} \left(
\frac{2}{3 \pi}\right)^{4/3} \frac{1}{\lambda^{1/3}} + \cdots \to 0 \,.
\ee
This says that the commutator is vanishing relative to observables
that only depend on single matrices. The ratio cancels out the
overall scaling of the $X$ and $Y$ matrices. Thus, the condition
that (\ref{eq:one}) go to zero seems to be a natural notion of
whether the matrices $X$ and $Y$ commute. As we discussed above,
another natural ratio to consider is
\be\label{eq:two}
\frac{\tr (XYXY)}{\tr (X^2 Y^2)} = 1 - \frac{35}{6 \pi}
\left(\frac{2}{3\pi}\right)^{1/3} \frac{1}{\lambda^{1/3}} + \cdots \to 1 \,.
\ee
Which also shows that the matrices commute. The parameter $\lambda^{1/3}$
controls the size of the non-commutativity of the matrices. If the
matrices are sufficiently close to commuting, we  make small errors by assuming that
they are mutually diagonal, and that there is a joint eigenvalue distribution.

Therefore, we consider the behaviour of the ratios (\ref{eq:one})
and (\ref{eq:two}) as criteria for the emergence of a local
geometry. Our computations above show that they are very closely
related to the mass of the off diagonal modes, $m^2_\text{o.d.}
\sim \lambda^{2/3}$, as we should expect.

One should notice that the criterion for commutativity, $r$ can be
refined further. For example, we can ask more local questions in
the spectrum of $X$ if we consider the ratios
\begin{equation}
r_f= \frac{\tr( f(X) Y f(X) Y)}{\tr( f(X)^2 Y^2)} \,,
\end{equation}
for $f$ some real function of $x$ which is peaked in some region.
We can do the same with $Y$. Depending on how $r_f$ varies with
the width $\delta$ of $f$, we can talk of a local degree of
noncommutativity on the scale of $\delta$. It should be noted that
these can be easily evaluated numerically if $f$ is a rational
function, without the need to diagonalize $X$. This  seems to
serve as a reasonable definition of the local sharpness of a fuzzy
geometry.

\subsection{Emergence of local geometry: hemisphere distribution}
\label{sec:emergence}

Given that most of the off diagonal modes of $Y$ are parametrically heavy
when $\lambda \gg 1$, and given the observations of the previous
subsection, we might expect to be able to recast this model as a
commuting matrix model for the two matrices $X$ and $Y$. A two
dimensional commuting matrix model is a matrix model in which the
matrices are further constrained to commute. At large $N$ the
model is described by a joint eigenvalue density $\rho(x,y)$ for
the eigenvalues of the commuting matrices.

\begin{figure}[h]
\begin{center}
\epsfig{file=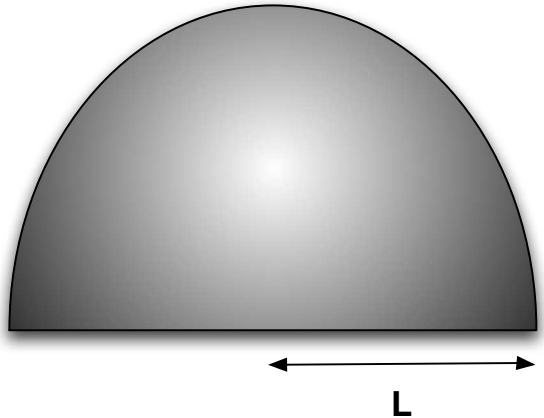,width=3in,angle=0,trim=0 0 0 0}%
\end{center}
\caption{The emergent two dimensional hemisphere distribution.
  \label{fig:hemis}}
\end{figure}

It is easy to see that the hemisphere distribution
\be\label{eq:2dsolution}
\rho(x,y) =
\left\{ \begin{array}{l}
\displaystyle \frac{3 N \sqrt{L^2 - x^2 - y^2}}{2 \pi L^3}
 \qquad \text{for} \quad x^2 + y^2 < L^2 \\
  0 \qquad \text{otherwise} \,, \\
\end{array} \right.
\ee
recovers our one dimensional parabolic distribution
(\ref{eq:solution}) upon integrating out one direction
\be\label{eq:integrateout}
\rho(x) = \int_{-\sqrt{L^2-x^2}}^{\sqrt{L^2-x^2}} \rho(x,y) dy \,.
\ee
This immediately implies that all observables depending on only
one of the matrices can be computed using this two dimensional
eigenvalue distribution, which we might call the hemisphere
distribution. Note that (\ref{eq:2dsolution}) is the unique
radially symmetric distribution with the property
(\ref{eq:integrateout}). Radial symmetry is appropriate as the
original two matrix model was $SO(2)$ invariant. The emergent
hemisphere distribution is shown in figure \ref{fig:hemis}.

A nontrivial test of this emergent two dimensional eigenvalue
distribution is to reproduce observables that depend on both
$X$ and $Y$. In particular we find
\be
\tr (XYXY) = \tr (X^2 Y^2) = \int \rho(x,y) x^2 y^2 dx dy = \frac{3 \pi}{70}
\left( \frac{3 \pi}{2} \right)^{1/3} \frac{N^3}{\lambda^{2/3}} \,.
\ee
Precisely reproducing the exact result (\ref{eq:fourpoint}) to
leading order at large $\lambda$. Here we note that off-diagonal
elements are manifestly not necessary in order to correctly compute
observables involving both X and Y in this commuting framework
(except for commutators, of course, which vanish to leading order).
Unlike in the previous discussion in which we diagonalised only
$X$, in this simultaneously diagonalised basis there is a
well-defined separation into light eigenvalues and heavy
off-diagonal modes.

It is natural to ask for an action that has the eigenvalue
distribution (\ref{eq:2dsolution}) as its ground state. The
na\"ive thing to do is to obtain a one loop effective action for
the simultaneous eigenvalues of $X$ and $Y$ by integrating out the
off diagonal modes. One might hope this will work even at strong
coupling, because, as we saw, the off diagonal modes are becoming
parametrically heavy. An estimate of when perturbation theory is
valid can be obtained as follows (essentially this argument
appears in, for instance, \cite{Hotta:1998en}). Given an
eigenvalue distribution of extension $L$, the action for off
diagonal modes $\delta X$ is schematically
\be
S_\text{off-diag} \sim (1 + g^2 L^2) \delta X^2 + g^2 L \delta X^3
+ g^2 \delta X^4 \,.
\ee
Supposing $g^2 L^2 \gg 1$, so that off-diagonal modes are heavy,
we can ask when the two loop contribution to the partition
function is parametrically smaller than the one loop partition
function. If we normalise the one loop contribution to 1, then it
is easy to see that the two loop contribution is of order $g^2
N/(g^2 L^2)^2$. Thus the higher loop contribution is negligible if
\be\label{eq:condition}
L \gg \frac{N^{1/2}}{\lambda^{1/4}} \,.
\ee

This condition is indeed satisfied by the width of the hemisphere
distribution (\ref{eq:L}). Emboldened, we go ahead and compute the
one loop action for the simultaneous eigenvalues to obtain
\be\label{eq:2daction}
S_{2D} = \sum_i \vec x_i^2 - \sum_{i \neq j} V(|\vec x_i - \vec
x_j|)
\,,
\ee
with
\be\label{eq:potential}
V(s) = \half \log(s^2) - \half \log (1+ g^2 s^2) \,.
\ee
In integrating out the off-diagonal modes one should gauge fix,
for instance as described in \cite{Berenstein:2005aa}. This is of
course the same action as we obtained before, except that now we
have two component vectors $\vec x = (x,y)$. The large $N$
equations of motion are
\be\label{eq:commuting}
\vec x = \int \frac{\vec x - \vec y}{|\vec x - \vec
y|} V'(|\vec x - \vec y|) \rho(\vec y) d^2y
\ee
Writing this equation out for a radially symmetric distribution
\bea
x & = & \int_0^L dr r \rho(r)
\int_0^{2\pi} d\theta \frac{V'(\sqrt{x^2 + r^2 - 2xr \cos\theta})
(x - r \cos \theta)}{\sqrt{x^2 + r^2 - 2xr \cos\theta}} \\
 & = & \frac{\pi}{x} \int_0^L dr r \rho(r)
 \left(\frac{1 + g^2 (r^2-x^2)}{\sqrt{1 + 2 g^2 (r^2 + x^2) + g^4 (r^2-x^2)^2}}
 + H(x-r) \right) \,.\label{eq:last}
\eea
In the second step we assumed, without loss of generality, that
$x>0$. We also introduced the step function $H(s)$ which equals
$-1$ for $s < 0$ and $+1$ otherwise.

At large $g L$ we can solve this last equation (\ref{eq:last})
with some educated guesswork. The solution is
\be
\rho(r) = \frac{2 N}{\pi} \frac{\widetilde L^2 - r^2}{\widetilde
L^4}\,.\label{eq:densityansatz}
\ee
where
\be\label{eq:log}
\widetilde L = N^{1/2} \left( \frac{\log \lambda}{\lambda}
\right)^{1/4}\,.
\ee
The correctness of this solution can be checked by performing the
integral (\ref{eq:last}) and then taking the strong coupling
limit. This clearly does not agree with the hemisphere
distribution (\ref{eq:2dsolution}) and the width $\widetilde L$
has a different scaling with $\lambda$ than (\ref{eq:L}). It is
interesting to see that the eigenvalue problem
(\ref{eq:commuting}) acquires a logarithmic non-analyticity in the
't Hooft coupling in the strong coupling limit. It is not easy to
check the solution (\ref{eq:log}) numerically to high accuracy,
because of the logarithmic dependence, but it easy numerically to
see the distribution become paraboloid and the width scale like
$1/\lambda^{1/4}$ rather than $1/\lambda^{1/6}$.

The implication of this mismatch is that the one loop effective
action for the eigenvalues (\ref{eq:2daction}) is insufficient to
capture the eigenvalue dynamics, despite the fact that the width
(\ref{eq:L}) satisfies the na\"ive bound (\ref{eq:condition}). We
suspect that this occurs because the scaling of $L
\sim 1/\lambda^{1/6}$ is fairly close to the limiting scaling
$1/\lambda^{1/4}$. It may be possible to identify the required
higher loop corrections to the effective action and re-obtain the
correct emergent geometry. Indeed the appearance of a logarithm of
the coupling suggests a resummation should be done, but we leave
this for future work.

To summarise this section: we have shown that the two matrix model
(\ref{eq:original}) becomes commuting at strong coupling and that
there is an emergent two dimensional hemisphere geometry
(\ref{eq:2dsolution}). We saw that locality of physics in this
geometry, i.e. that the off diagonal modes are heavy, was captured
by the observables (\ref{eq:one}) and (\ref{eq:two}). We found
that describing the eigenvalue dynamics will require going beyond
the one loop effective action.

\section{Solvable models with more than two matrices}

\subsection{Bosonic model: No geometry}
\label{sec:boson}

There is a generalisation of the two matrix model that can also be
treated analytically. It is instructive to see how this model does
not lead to an emergent geometry at strong coupling. Consider the
$k+1$ matrix model with $SO(k)$ symmetry only
\be\label{eq:Zk}
Z = \int \calD X \calD Y_1 \cdots \calD Y_k e^{- \tr X^2 -
\sum_m \tr Y^2_m + g^2 \sum_m \tr [X,Y_m]^2}\,.
\ee
This integral is again quadratic in all the $Y_m$s, which may
therefore be integrated out exactly. The crucial simplification is
the absence of interactions between the $Y$ matrices.
Diagonalising $X$, we obtain
\be\label{eq:kpart}
Z = \int dx_1 \ldots dx_N e^{-
\sum_i x_i^2 + \frac{1}{2} \sum_{i \neq j} \log(x_i-x_j)^2 - \frac{k}{2} \sum_{i \neq j} \log[1 +
g^2 (x_i-x_j)^2]}\,.
\ee
The large $N$ saddle point of this integral is an eigenvalue
distribution satisfying
\be\label{eq:kintegral}
x = \int \left[\frac{1-k}{x-y} +  \frac{k}{(x-y) [1 + g^2
(x-y)^2]} \right] \rho(y) dy \,.
\ee
We have rearranged the terms a little. As before $\int \rho(x) dx
= N$. The second term is a repulsive force of the same form as we
found previously in (\ref{eq:integral}). The first term, however,
is now an attractive force (for $k>1$) that is stronger at long
distances. Eigenvalues separated by $(x-y)^2 = 1/(g^2 (k-1))$
experience no net force. One could presumably solve this integral
equation fully using the techniques in \cite{Kazakov:1998ji}. Once
again, we shall look for a pedestrian approach using the strong
coupling expansion.

A good starting point to attack (\ref{eq:Zk}) analytically is the
limit of a large number $k$ of matrices. The attractive force is
becoming more important in this limit, so we might expect the
width $L$ of the distribution becomes small. In fact, based on the
force balance equation, the natural scale to expect is $L^2 \sim
1/g^2 k$. Let us assume this scaling and see where it leads.

If $L^2 \sim 1/g^2 k$ and $k$ is becoming large, then $(x-y)^2 g^2
\leq 4 L^2 g^2$ is becoming small. Therefore we can expand the
last term in (\ref{eq:kintegral}). Rearranging, we obtain
\be\label{eq:kintegral2}
x + k g^2 \sum_{n=0}^{\infty} (-1)^n g^{2n} \int (x-y)^{2n+1}
\rho(y) dy = \int \frac{\rho(y) dy}{x-y}  \,.
\ee
The first few terms in the sum read
\be\label{eq:eqpand}
(1+k \lambda) x - \frac{k \lambda^2}{N} (x^3 + 3 \a_2 x) + \frac{k
\lambda^3}{N^2} (x^5 + 10 \a_2 x^3 + 5 \a_4 x) + \cdots = \int
\frac{\rho(y) dy}{x-y}  \,.
\ee
We have used the fact that $\rho(y)$ will be even and introduced
the notation for the moments of the distribution
\be
\a_m = \frac{1}{N} \int y^m \rho(y) dy \,.
\ee
Note that each term in this expansion is suppressed by a factor of
$L^2 g^2 \sim 1/k$ compared to the previous terms. Therefore we
can solve order by order in the large $k$ limit. The leading order
solution is clearly the semicircle
\be\label{eq:firstorder}
\rho(x) = \frac{2 N}{\pi L^2} \sqrt{L^2 - x^2} \,,
\ee
with
\be\label{eq:Lk}
L = \sqrt{\frac{2 N}{1+k\lambda}} \,.
\ee
This expression implies $L^2 g^2 \ll 1$ for all $\lambda$, in the
large $k$ limit. Therefore this is the correct leading order
solution at large $k$ for all couplings, not just strong coupling.
It is not surprising that upon integrating out a large number of
matrices one obtains the semicircle distribution, corresponding to
effectively Gaussian degrees of freedom. We have checked that
(\ref{eq:Lk}) agrees excellently with numerical results.

\begin{figure}[h]
\begin{center}
\epsfig{file=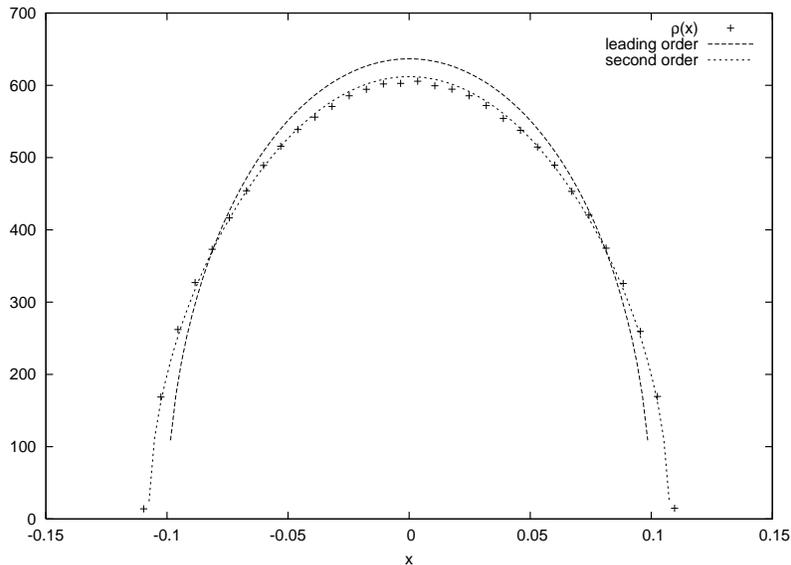,width=3in,angle=270,trim=0 0 0 0}%
\end{center}
\caption{Numerically simulated eigenvalue distribution solution to
(\ref{eq:kpart}) with $N=100$, $k=20$ and $\lambda = 1000$,
together with the theoretical result to first
(\ref{eq:firstorder}) and second (\ref{eq:rhosecond}) order in a $1/k$ expansion.
The leading order distribution has $L=0.0975$ and the subleading
distribution $L=0.1075$.
  \label{fig:order2}}
\end{figure}

To second order, the solution may be found using standard matrix
model techniques. It is straightforward to verify that the solution to second order is
\be\label{eq:rhosecond}
\rho(x)  =  \frac{N}{\pi L^2} \sqrt{L^2 - x^2} \left(2 + \frac{k \l^2 L^4}{4 N^2} - \frac{k \l^2 L^2}{N^2} x^2 \right)
\,,
\ee
where the width $L$ satisfies (to this order in large $k$)
\be
4 N^2 - 2 N L^2 (1+ k \l) + 3 k L^4 \l^2 = 0 \,.
\ee
If we work in the large $\lambda$ limit for simplicity, this is
seen to imply a corrected eigenvalue width of
\be
L = \sqrt{\frac{2 N}{k \lambda}} \left(1 + \frac{3}{2 k} \right)
\,.
\ee
We checked these results by simulating the eigenvalue partition
function (\ref{eq:kpart}) numerically. We again used Hybrid
Monte-Carlo with $N=100$ and stored 1000 configurations,
determining the distribution from $N \times 1000$ points. The
resulting distribution for $\lambda = 1000$ and $k=20$ is shown in
figure
\ref{fig:order2}  together with the theoretical expectation
(\ref{eq:rhosecond}). There is an excellent agreement.

It is straightforward to move systematically to arbitrary order. For instance the seventh order
solution is of the form
\be\label{eq:rhoexpand}
\rho(x)  =  \frac{N}{\pi} \sqrt{L^2 - x^2} \left(a + b x^2 + c x^4 + d x^6 + e x^8 + f x^{10} + g x^{12} \right)
\,,
\ee
with the constants $a, \cdots , g$ determined by plugging this
expression into (\ref{eq:kintegral2}). The width of the
distribution (at large $\lambda$) is found to be at this order
\be\label{eq:Lexpand}
L = \sqrt{\frac{2 N}{\l k}} \left(1 + \frac{3}{2} \frac{1}{k} +
\frac{5}{4} \frac{1}{k^2} + \frac{11}{8} \frac{1}{k^3} +
\frac{57}{32} \frac{1}{k^4} + \frac{33}{64} \frac{1}{k^5} +
\frac{39}{16} \frac{1}{k^6}\right) \,.
\ee
For $k=2$ this formula gives $L =  2.3999 \sqrt{N/\l}$ with small
corrections. Given that $k=2$ is the smallest case we have to
consider, this implies that (\ref{eq:rhoexpand}) and
(\ref{eq:Lexpand}) are a good approximation to the solution for
all integer $k > 1$. The solution has $L^2 \sim 1/g^2k$, and so
selfconsistently satisfies our scaling assumption. It seems
plausible that the radius of convergence of the series
(\ref{eq:Lexpand}) will be $k=1$, consistent with the fact that we
found a different scaling with $\l$ in that case.

The implication of the above results, e.g. (\ref{eq:Lexpand}), is
that for these multi-matrix models, with $k \geq 2$, the mass of
the off diagonal modes of the matrix $X$ is of the same order as
the diagonal modes. This is because $L^2 g^2$ is order one or
smaller. Therefore we do not expect an effective description in
terms of the simultaneous eigenvalues of $X$ and $Y$. Furthermore,
we don't expect the $Y$s to commute amongst themselves as they
only couple to one another through $X$. Let us see if these
expectations are reflected in the observables we considered
previously.

For simplicity, let us work to leading order at large $\lambda$
and $k$. That is, we take the solution (\ref{eq:firstorder}) for
$\rho(x)$. We have seen that the scaling of quantities with
$\lambda$ does not change away from this limit down to $k=2$.

The two point function of $X$ is easy to evaluate
\be
\frac{\tr X^2}{N} = \frac{1}{N} \int_{-L}^L x^2 \rho(x) dx =
\frac{L^2}{4} = \frac{N}{2 k \l} \,.
\ee
To compute correlation functions involving the $Y$s, we introduce
source terms as before
\be
Z[J] = \int \calD X \calD Y_1 \cdots \calD Y_k e^{- \tr X^2 -
\sum_m \tr Y^2_m + g^2 \sum_m \tr [X,Y_m]^2 + \sum_m \tr J_m Y_m}
\,.
\ee
Using exactly the same steps as in section \ref{sec:emergence}
above, we can obtain the two point function for the $Y$s
\be
\frac{\tr Y_n^2}{N} = \frac{1}{N} \int \frac{\rho(x) \rho(y) dx dy}{2
(1+g^2(x-y)^2)} \approx \frac{1}{2 N} \int \rho(x) \rho(y) dx dy =
\frac{N}{2} \,.
\ee
In this and subsequent formulae we are using the fact noted above
that $(x-y)^2 g^2 \leq 4 L^2 g^2 \ll 1$ for this distribution. We
see that the $Y$s are spread out by an extra factor of $\lambda$
compared to the $X$ matrix. The four point functions $\tr X^2 Y^2$
and $\tr XYXY$ are
\bea
\tr X^2 Y_n^2 & = & \int \frac{\rho(x) \rho(y) x^2 dx dy}{2
(1+g^2(x-y)^2)} \nonumber \\
& \approx & \frac{1}{2} \int \rho(x) \rho(y) x^2 dx dy = \frac{L^2
N^2}{8} = \frac{N^3}{4 k \l}
\,,
\eea
and
\bea
\tr X Y_n X Y_n & = & \int \frac{\rho(x) \rho(y) xy dx dy}{2
(1+g^2(x-y)^2)} \nonumber \\
& \approx & - \frac{g^2}{2} \int (x-y)^2 x y \rho(x)
\rho(y) dx dy = \frac{\l L^4 N}{16} =
\frac{N^3}{4 k
\l} \frac{1}{k} \,.
\eea
This second expression is suppressed by an extra factor of $1/k$
and so can be neglected to leading order at large $k$. It follows
that to leading order $\tr [X,Y]^2 = - 2 \tr X^2 Y^2$.

The four point functions involving two different $Y$s are
\be
\tr Y_m^2 Y_n^2 = \int \frac{\rho(x) \rho(y) \rho(z) dx dy
dz}{4 (1+g^2(x-z)^2) (1+g^2(y-z)^2)} \approx \frac{1}{4} \int
\rho(x) \rho(y) \rho(z) dx dy dz = \frac{N^3}{4} \,,
\ee
and
\be
\tr Y_m Y_n Y_m Y_n = \frac{N}{4} \sim 0 \,.
\ee
This last term is subleading in $1/N$ and therefore effectively
zero in the planar approximation we are taking. The following four
point functions are also obviously zero at large $N$, from the
absence of $Y$ mixing in the classical action
\be
\tr X Y_m X Y_n = \tr X^2 Y_m Y_n =  0 \,.
\ee
We have checked all the four point functions in this section
by comparing with numerical results obtained by simulating
the full partition function (\ref{eq:Zk}).

We can now compute our ratios to be
\be
\frac{N \tr [X,Y_n]^2}{\tr X^2 \tr Y_n^2} = \frac{N \tr [Y_m,Y_n]^2}{\tr Y_m^2 \tr Y_n^2} = -2 \,,
\ee
and
\be
\frac{\tr X Y_n X Y_n}{\tr X^2 Y_n^2} = \frac{1}{k} \to 0 \,.
\ee
Both of these expressions are consistent with our expectation that
this model is not commuting. The fact that the first ratio is not
going to zero and the second is less than one will remain true at
finite $k$. It is interesting to note that at finite $k$ the
second ratio is not driven to zero as the coupling goes to
infinity. Therefore the finite $k$ models are not `maximally
non-commuting'. For the $Y$s however, we do have at large $N$ that
\be
\frac{\tr Y_m Y_n Y_m Y_n}{\tr Y_m^2 Y_n^2} = 0 \,.
\ee
A simple but important lesson to draw from this model is that
quantum loop effects are crucial in determining whether the
strongly coupled system is commuting or not. Here, the extra one
loop contributions from the matrices destabilised the emergent
geometry of the two matrix model. This suggests that we can
improve the situation by adding fermionic matrices to cancel the
undesired loop contributions.

A final point to note, from e.g. (\ref{eq:eqpand}), is that the
mass term is unimportant in determining the strong coupling
eigenvalue distribution in this model. This is in contrast to the
(commuting at strong coupling) two matrix case we considered
previously. Without a mass term, the coupling $\lambda$ can simply
be absorbed into the normalisation of the $X$ matrix and therefore
does not have any dynamics associated to it. In particular, strong
coupling cannot drive us to commutativity in a massless theory. A
lesson to draw, therefore, is that commutativity should involve an
interplay been the mass term and the `commutator square'
interaction terms.

\subsection{Model with fermionic matrices: still no geometry}
\label{sec:fermion}

It has long been appreciated that supersymmetry facilitates the
emergence of geometry \cite{Banks:1996vh, Ishibashi:1996xs,
Dijkgraaf:1997vv}. Here we show how a simple implementation of
this idea works for us. However, the absence of direct
interactions between the $Y$ matrices will ultimately prevent the
emergence of a geometry involving all the matrices in these
solvable models.

We can supplement the $1+k$ matrix model (\ref{eq:Zk}) with $2 h$
fermionic fields as follows
\bea\label{eq:Zfermions}
Z & = & \int \calD X \calD Y_1 \cdots \calD Y_k \calD \l_1 \cdots
\calD \l_h \calD \mu_1 \cdots \calD \mu_h \nonumber \\
& & \qquad  e^{- \tr X^2 - \sum_m \tr Y^2_m - \sum_n \tr \l_n
\mu_n + g^2
\sum_m \tr [X,Y_m]^2 + i g \sum_n \tr \l_n [X,\mu_n]}\,.
\eea
In this model, the $\l_n$ and $\mu_n$ are Hermitian $N$ by $N$
matrices of anticommuting numbers. Each matrix component of $\l$
and $\mu$ has only a single component, no `spinor index' is
necessary. Because of the asymmetry of the interactions, we are
able to add fermions without introducing a Clifford algebra and
without breaking any of the symmetries of the bosonic action.

The partition function is quadratic in all the $Y$s and fermions,
which we integrate out exactly to give
\be
Z = \int dx_1 \ldots dx_N e^{-
\sum_i x_i^2 + \frac{1}{2} \sum_{i \neq j} \log(x_i-x_j)^2 + \frac{h-k}{2} \sum_{i \neq j} \log[1 +
g^2 (x_i-x_j)^2]}\,.
\ee
We see that including the fermions can cancel out the attractive
part of the potential, thus increasing the likelihood that the
eigenvalues of $X$ will spread out sufficiently to give the off
diagonal modes with the $Y$s a large mass. We already know the
answer for the eigenvalue distribution in several cases:
\begin{itemize}
\item If $h < k-1$, the partition function is (\ref{eq:Zk}) with
$k' = k-h > 1$ and one obtains the same results as in the previous
subsection, with no emergent geometry.

\item If $h=k-1$, the partition function becomes precisely that of
section \ref{sec:parabolic}, leading to a parabolic distribution
of the eigenvalues of $X$ with width $L \sim g^{-1/3}$ at strong
coupling.

\item For the case $h=k$, the partition function is that for a Gaussian
model and clearly results in a semi-circle distribution for the
eigenvalues of $X$ with width $L \sim N^{1/2}$.

\end{itemize}
In the latter two cases, $g^2 L^2$ becomes large at strong
coupling and therefore the off diagonal modes between $X$ and any
of the $Y$s become heavy. Thus they are candidates for emergent
geometry. However, we will shortly see that the $Y$s will not
commute amongst themselves. This is essentially an artifact of the
simple action we have taken, with no interactions between the
$Y$s. Generically, if two Hermitian matrices $Y_m$ and $Y_n$ both
commute with $X$ then they should commute amongst themselves. The
exception is if $X$ has degenerate eigenvalues. This is
effectively what is happening here.

For the remaining cases, with $h>k$, all the terms in the
effective action lead to repulsive forces. It is useful to write
the large $N$ equations of motion as
\be\label{eq:kheqns}
x =  \int \left[\frac{1+h-k}{x-y} +  \frac{k-h}{(x-y) [1 + g^2
(x-y)^2]} \right] \rho(y) dy \,.
\ee
Because the force is more repulsive than the case $h=k$, we expect
that the distribution $\rho(x)$ will have width $L \gtrsim
N^{1/2}$. This would imply that $g L \gg 1$ at strong coupling.
Assuming this, then from (\ref{eq:general}) we can ignore the last
term in (\ref{eq:kheqns}). It follows that the distribution at
strong coupling is a semicircle
\be\label{eq:largelambda}
\rho(x) = \frac{2 N}{\pi L^2} \sqrt{L^2 - x^2} \,,
\ee
with width
\be\label{eq:Llargelambda}
L = \sqrt{2 N (1+h-k)} \,.
\ee
This result includes the case $h=k$. The result self-consistently
satisfies $g L \gg 1$ in the strong 't Hooft coupling limit.
Therefore we can conclude that for all $h \geq k - 1$, the off
diagonal modes of the $Y$ matrices relative to $X$ become heavy.

It follows that
\be\label{eq:XX}
\frac{\tr X^2}{N} = \frac{L^2}{4} = \frac{N (1+h-k)}{2} \,,
\ee
and, to leading order at large $\lambda$ using for instance
(\ref{eq:general}) to compute the large $\lambda$ limit,
\be\label{eq:YY}
\frac{\tr Y^2_n}{N} = \frac{8 N}{3 \pi g L} = \frac{4 \sqrt{2} N}{3 \pi
(1+h-k)^{1/2}} \frac{1}{\l^{1/2}}\,.
\ee
The commutator can be computed robustly as we did for the two
matrix case, by differentiating the logarithm of the partition
function with respect to $\lambda$. To pick out the bosonic
commutator squared, one should take the couplings in front of the
bosonic and fermionic interactions to be distinct before
differentiating. We find
\be\label{eq:commutatorfermions}
\tr [X,Y_n]^2 = - \frac{N^3}{2 \lambda} \,.
\ee
Also proceeding as previously, we find the four point functions
\bea
\tr X^2 Y_n^2 & = & \frac{8 L N^2}{15 \pi g} - \frac{N^2}{4 g^2} + \frac{N^2 \b}{g^2} \nonumber \\
& = & \frac{8 \sqrt{2} (1+h-k)^{1/2} N^3}{15 \pi}
\frac{1}{\l^{1/2}} - \frac{N^3}{4 \l} + \frac{N^3 \b}{\l} \,,
\\
\tr X Y_n X Y_n & = & \frac{8 L N^2}{15 \pi g} - \frac{N^2}{2 g^2}  + \frac{N^2 \b}{g^2} \nonumber \\
& = &
\frac{8 \sqrt{2} (1+h-k)^{1/2} N^3}{15 \pi} \frac{1}{\l^{1/2}} - \frac{N^3}{2 \l} + \frac{N^3 \b}{\l} \,, \\
\tr Y_m^2 Y_n^2 & = & \frac{3 N^3}{4 g^2 L^2} = \frac{3 N^3}{8
(1+h-k)} \frac{1}{\l} \,, \\
\tr Y_m Y_n Y_m Y_n & = & \frac{N}{4} \sim 0 \,.
\eea
As before, there is an unknown subleading contribution $N^3 \b/\l$
due to the leading correction to the eigenvalue distribution at
large $\lambda$. We know that the contribution is equal in both
the expressions above, because the difference has to reproduce the
commutator (\ref{eq:commutatorfermions}). We do not need the value
of $\b$ for our computations below, and we do not attempt to
calculate it.

The above results hold for $h \geq k$. In the `critical' case $h =
k-1$, the four point correlator of the $Y$s was not computed
previously. Using the distribution (\ref{eq:solution}) we find to
leading order at strong coupling
\be
\tr Y_m^2 Y_n^2 = \frac{27 \pi^2 N^3}{280 g^2 L^2} = \frac{9 \pi
N^3}{140}\left(\frac{3 \pi}{2} \right)^{1/3} \frac{1}{\l^{2/3}}
\,.
\ee

We can now discuss the emergence of geometry in the models with $h
\geq k - 1$, that is, with sufficiently many fermions that the off
diagonal modes of $X$ with any of the $Y_m$ matrices are heavy at
strong coupling $\lambda \to \infty$. Two comments apply to all of
these cases. Firstly, as we should expect, the $X$ matrix commutes
with the $Y_m$ matrices according to both of our criteria. Namely
\be
\frac{N \tr [X,Y_m]^2}{\tr X^2 \tr Y_m^2} \to 0 \,, \qquad \frac{\tr X Y_m X Y_m}{\tr
X^2 Y_m^2} \to 1 \,, \qquad \text{as} \qquad \lambda \to \infty
\,.
\ee
This is easily seen from our above expressions for the relevant
two and four point functions.

Secondly, although one might therefore have expected the $Y$s to
commute amongst themselves, this is not the case. For all of these
models it follows from our above expressions that
\be
\frac{N \tr [Y_m,Y_n]^2}{\tr Y_m^2 \tr Y_n^2} \to {\mathcal{O}}(1) \,, \qquad \frac{\tr Y_m Y_n Y_m Y_n}{\tr
Y_m^2 Y_n^2} \sim 0 \,, \qquad \text{as} \qquad \lambda \to \infty
\,.
\ee
The fact that the $Y$s do not commute amongst themselves can be
understood as being due to the absence of interactions between the
$Y$ matrices in the action of (\ref{eq:Zfermions}). At leading
order in large $N$ they are simply uncorrelated and cannot
commute. Unfortunately, the very simplification that allowed us to
get an analytic handle on this model undoes the possibility of an
emergent geometry in which all of the matrices participate. The
lesson we might take away is that genericity is another important
property in the search for commutating models: if a matrix has
degenerate eigenvalues it can commute with two other matrices
without the other matrices needing to commute amongst themselves.

Given that the $X$ matrix commutes with all of the $Y$ matrices,
one might imagine picking one of the $Y$ matrices, say $Y_1$ (or
perhaps some combination of them, to preserve the $SO(k)$
invariance) and considering a geometry given by the joint
eigenvalue distribution of this matrix with $X$. However, this
geometry is not especially useful, even in the critical case $h =
k-1$, where the $X$ and $Y_1$ eigenvalues would have roughly the
same spread.\footnote{The cases with $h \geq k$ are even worse.
The spread of the $Y_1$ eigenvalues is smaller by a power of
$\lambda$ compared to the spread of the $X$ eigenvalues, see
(\ref{eq:XX}) and (\ref{eq:YY}). In commuting matrix models, a large
anisotropy can prevent the smaller dimension from emerging at all
\cite{Gursoy:2007np, Aharony:2007rj}.} The problem is that the physics
of the remaining $Y$ matrices would not be local in the $y_1$
direction. Off diagonal modes of $Y_2$ (say) connecting different
values of $y_1$ would have a comparable effect on the dynamics as
the eigenvalues of $Y_2$. So the strongly coupled matrix model is
not solved by local two dimensional geometric physics in these
cases. For emergent local geometry, {\emph{all}} modes relating
far away spacetime points should be massive compared to local
modes.

The natural next step, in search for higher dimensional geometry,
is to introduce interactions between the $Y$s. This substantially
increases the difficulty of solving the model analytically.

\section{The fully interacting multi-matrix model}

\subsection{Bosonic model}

The interacting $p$-matrix bosonic model with full $SO(p)$
invariance is
\be\label{eq:fullbosonic}
Z = \int \calD X_1 \cdots \calD X_p  e^{- \sum_m \tr X_m^2 +
\frac{1}{2} g^2 \sum_{m,n} \tr [X_m,X_n]^2}\,.
\ee
We cannot solve this model exactly, for $p>2$, even in a strong coupling expansion.
However, building on the intuition from previous cases, a few observations
are possible.

We do not expect this bosonic model to be commuting. We saw above that
multiple one loop bosonic contributions result in an attractive potential at long
distances. The eigenvalues were therefore insufficiently spread out for the
off-diagonal modes to become parametrically heavier than the eigenvalues at
strong coupling. Thus we do not expect an emergent geometry. We will
study a commuting ansatz for a similar matrix model shortly. Let us suppose
for the moment that all the elements of the matrices, diagonal and off diagonal,
are of the same order. By $SO(p)$ invariance, the entries will be of the same
magnitude for all the matrices
\be
\langle X_{ij} \rangle  \sim \frac{1}{m_\text{eff.}} \,.
\ee
Here $m_\text{eff.}$ is the effective mass of the matrix elements. If the
matrices become heavy at strong 't Hooft coupling, then we might hope to
trust a one loop evaluation of their masses
\be\label{eq:consistent}
m_\text{eff.}^2 \sim 1 + g^2 \sum_k \langle X_{ik} X_{kj}  \rangle
\sim \frac{\lambda}{m_\text{eff.}^2} \,.
\ee
From which we would conclude that the typical matrix entry scales
as
\be\label{eq:scaling}
\langle X_{ij} \rangle \sim \frac{1}{\lambda^{1/4}} \,.
\ee
Note that if we diagonalise one of the $X$ matrices, with entries
of order (\ref{eq:scaling}), we will obtain diagonal elements of
order $N^{1/2}/\lambda^{1/4}$. This is seen, for instance, from
the observation that (\ref{eq:scaling}) implies $\tr X^2 = X_{ij}
X_{ji} \sim N^2/\lambda^{1/2}$ whereas for a diagonal matrix $\tr
X^2 = X_{ii}^2$. Thus (\ref{eq:scaling}) is precisely on the
boundary of our condition (\ref{eq:condition}) for trusting a one
loop effective action. We will shortly give an independent
argument supporting the scaling (\ref{eq:scaling}).

In the previous paragraph, using selfconsistency and a
non-commutativity assumption, we obtained a scaling for the spread
of matrix elements with $\lambda$ at strong coupling. The $SO(p)$
invariance of the model was also important; in section
\ref{sec:boson} above an anisotropic model gave different scalings.

A consistency check of the above picture is that we can compute
one of our ratios, following e.g. \cite{Hotta:1998en}. Consider
the change of variables $X_m \to (1 +
\epsilon) X_m$. This must leave the partition function invariant.
To linear order the measure for each matrix changes as $\calD X_m
\to (1 + (N^2-1) \epsilon) \calD X_m$. Requiring the partition function to be invariant
and using the $SO(p)$ symmetry leads to
\be\label{eq:niceformula}
p (N^2-1) = 2 p \langle \tr X_m^2 \rangle - 2 g^2 p (p-1)
\langle \tr [X_m, X_n]^2 \rangle \,.
\ee
This is an exact expression. Now using the scaling
(\ref{eq:scaling}) at large coupling and taking the large $N$
limit we find
\be\label{eq:comobser}
\frac{N \tr [X_m, X_n]^2}{\tr X_m^2 \tr X_n^2}
= \frac{- N^3}{2 (p-1) g^2 (\tr X_m^2)^2} \sim {\mathcal{O}}(1)
\,.
\ee
Thus, consistently, the matrices are indeed not commuting.

At this point we can note a general result for models with these
types of interactions. The model will be commuting, according to
the observable (\ref{eq:comobser}), if and only if the eigenvalues
are more spread out than $N^{1/2} \lambda^{-1/4}$ as $\lambda \to
\infty$. Indeed, so far the only model we have discussed that
satisfied this property for all the matrices involved was the two
matrix model. The property is precisely the same condition that we
found for the one loop effective action to be reliable. Although,
as we saw in the two matrix case, this latter condition is not
precise.

Consistently with our discussion at the end of section
\ref{sec:boson} above for non-commuting models, we see that
the mass term is not playing a role in any of our strong coupling
considerations. See for instance (\ref{eq:consistent}) or
(\ref{eq:niceformula}). This allows us to make contact with
previous numerical results on the model (\ref{eq:fullbosonic}) in
the absence of a mass term \cite{Krauth:1998yu, Krauth:1999qw,
Hotta:1998en}.

If we na\"ively take the strong coupling limit of the action
(\ref{eq:fullbosonic}), then we might expect to be able to drop
the mass term. In the massless theory we could eliminate the
coupling by rescaling the matrices $X \to X \lambda^{-1/4}$. It is
then immediate that the entries of $X$ will scale like $X_{ij}
\sim
\lambda^{-1/4}$ and therefore that eigenvalues will scale like $x
\sim N^{1/2}
\lambda^{-1/4}$. This scaling was indeed found numerically in
\cite{Hotta:1998en} for $p > 2$. These results support our
discussion above, including the observation that the mass term
appears to indeed be unimportant at strong coupling in these
models. For $p=2$ the mass term is important because there is an
infrared divergence in the massless model \cite{Krauth:1998yu,
Krauth:1999qw} due to the zero modes in the commutator squared
potential. This is consistent with the fact that we found a
different scaling in the two matrix model case.

The bottom line for these fully interacting bosonic models with
$p>2$ appears to be that the strong coupling physics essentially
reduces us to the massless case. The scaling of quantities with
$\lambda$ is fixed by dimensional analysis and the model is not
commuting. Diagonal and off-diagonal modes contribute equally to
generic observables, so there is no emergent local geometry.
Clearly the arguments in this subsection are not intended to be as
rigorous as our previous considerations.

\subsection{`Supersymmetrised' models}

We found in a solvable model above that adding fermions to the
model such that the fermion determinant cancelled the bosonic one
loop contribution allowed the eigenvalues to spread out.
Supersymmetrisations of the massless fully interacting bosonic
model exist in dimensions $p=3,4,6,10$, corresponding to
${\mathcal{N}} = 2,4,8,16$ real supercharges, respectively. These
have been studied numerically in for instance
\cite{Krauth:1999qw, Ambjorn:2000bf}. In the cases of $p=6$ and
$p=10$, the sign problem of the fermion determinants means that
thinking in terms of a positive eigenvalue distribution is
potentially misleading. However, in the cases of $p=3$ and $p=4$
there is no sign problem at large $N$ \cite{Krauth:1998xh,
Ambjorn:2000bf}. Furthermore, the (massless) cases $p=3$ and $p=4$
show strong infrared effects: the partition function is divergent
for $p=3$ whereas for $p=4$ the partition function is finite but
all moments $\langle \tr X^{2m}
\rangle$ diverge \cite{Krauth:1999qw}. Therefore these models are excellent
candidates for an emergent commuting geometry upon adding a mass
term, as we would expect a balance between the repulsive
interactions and the confining mass term.

Adding a supersymmetric mass term to the massless models is not
straightforward, however. Instead we will consider simpler models.
We start with the following $SO(p)$ invariant matrix model,
defined for all $p$, with $2(p-1)$ Hermitian fermionic matrices
$\lambda_m$ and $\mu_m$
\bea\label{eq:fullfermions}
Z & = & \int \calD X_1 \cdots \calD X_p \calD \l_1 \cdots
\calD \l_{p-1} \calD \mu_1 \cdots \calD \mu_{p-1} \nonumber \\
& & \qquad  e^{- \sum_m \tr X^2_m + \frac{1}{2} g^2 \sum_{m,n}
\tr [X_m,X_n]^2 - \sum_m \lambda_m \sqrt{1 + \frac{1}{2} g^2 \sum_n [X_n, \, \bullet]^2} \; \mu_m}\,.
\eea
The fermionic interaction has been chosen so that the bosonic and
fermionic one loop determinants cancel. Note that $2(p-1)$ is the appropriate
number of massive fermions to cancel the massive bosons,
because $X_m \to U X_m U^{-1}$ is a massless mode.
In integrating out the off diagonal modes, one should for
instance gauge fix as described in \cite{Berenstein:2005aa, Berenstein:2005jq}.

Let us assume that this model is commuting at strong coupling and
see if we can find a self consistent solution. If the off diagonal
modes are heavy about this commuting background, we can hope that
a one loop integration is sufficient. The eigenvalues are now
vectors in $\R^p$ given by $\vec x_i = (x^1_i, \cdots, x^p_i)$.
The partition function becomes
\be\label{eq:part}
Z = \int d\vec x_1 \ldots d\vec x_N e^{-
\sum_i \vec x_i^2 + \sum_{i \neq j} \frac{1}{2} \log|\vec x_i- \vec x_j|^2}\,.
\ee
The $\log|\vec x_i- \vec x_j|^2$ term can be thought of as a
generalised Vandermonde determinant arising from simultaneously
diagonalising the $p$ matrices. The large $N$ equations of motion
are
\be\label{eq:psusy}
\vec x =  \int d^py \rho(\vec y) \frac{\vec x- \vec y}{|\vec x- \vec y|^2} \,.
\ee
These equations of motion have been considered in some detail in,
for instance, \cite{Berenstein:2005aa, Berenstein:2005jq,
Aharony:2007rj}. The solution is given by a $(p-1)$ sphere with
constant eigenvalue density
\be\label{eq:sphere}
\rho(\vec y) = \frac{N}{|\vec y|^{p-1} \text{Vol} S^{p-1}} \,
\delta\left(|\vec y|^2 - \frac{N}{2} \right)\,.
\ee
From the spherical eigenvalue distribution (\ref{eq:sphere}) one
immediately obtains
\be
\frac{\tr X_m^2}{N} = \frac{N}{2p} \,.
\ee
At strong coupling, the width of the eigenvalue distribution
(\ref{eq:sphere}) is well within our bound (\ref{eq:condition})
for when the one loop effective potential is expected to be valid.
Therefore it seems likely that we have found a consistent saddle
point for the full model. Without having solved the model exactly
we cannot prove that this commuting saddle is the dominant large
$N$ saddle of the integral. However, we can see that
(\ref{eq:psusy}) describes a balance between a mass term and an
eigenvalue repulsion of the type we found previously in a
commuting model. The eigenvalues have spread out as far as the
mass term allows and simultaneously made the off diagonal modes
parametrically heavy at strong coupling. Therefore, there would
seem to be a good chance that this model is indeed showing an
emergent geometry at strong coupling.

The eigenvalue distribution (\ref{eq:sphere}) together with the
validity of the one loop effective action allows us to compute the
commutator
\bea
\tr [X_m, X_n]^2 & = & - \frac{1}{p} \int \frac{|\vec y - \vec y'|^2 \rho(\vec y) \rho(\vec y')
d^py d^p y'}{1 + g^2 |\vec y - \vec y'|^2} \nonumber \\
& = & - \frac{N^3 \text{Vol} S^{p-2}}{p \text{Vol} S^{p-1}}
\int_0^\pi \frac{(\sin\theta)^{p-2} (1 - \cos\theta)}{1 + \lambda (1-
\cos\theta)}d\theta \nonumber \\
& \to & - \frac{N^3}{p} \frac{1}{\lambda} \,, \qquad
\text{as}
\qquad
\lambda \to \infty \,.
\eea
Here we used the fact that the sphere has radius $r^2 = N/2$. We
can also compute the leading order four point functions
\be
\tr X_m^2 X_n^2 = \tr X_m X_n X_m X_n = \int d^py \rho(\vec y)
y_m^2 y_n^2 = \frac{r^4}{p(p-1)} - \frac{\langle y_m^4
\rangle_{S^{p-1}}}{p-1} = \frac{N^3}{4 p (p+2)}
\,.
\ee
This computation simply involves integrations over the $(p-1)$
sphere. It should be possible to test these commuting saddle
results using a numerical simulation of the full partition
function (\ref{eq:fullfermions}).

Other models with $SO(p)$ invariance that have an a priori chance
of having commuting saddles are the massless supersymmetric models
we described at the start of this section together with a mass
term for the bosons only:
\be\label{eq:Zsusymass}
Z = \int \calD X \calD \Psi e^{- \sum_m \tr X_m^2 + \frac{1}{2}
g^2 \sum_{m,n} \tr [X_m, X_n]^2 + g \sum_{m,\a,\b}\tr \Psi_\a
[\Gamma_{\a\b}^m X_m, \Psi_\b]} \,.
\ee
We can use the gamma matrix conventions of for instance
\cite{Krauth:1998xh}. In any case all that is important for us is
that in the massless case the one loop bosonic and fermionic
determinants cancel about a commuting background. We ignore for
the moment the fermion zero modes. Therefore in our model, the one
loop effective action about a commuting background is
\be\label{eq:susymass}
Z = \int d\vec x_1 \ldots d\vec x_N e^{-
\sum_i \vec x_i^2 + \frac{p-1}{2} \sum_{i \neq j} \left[ \log |\vec x_i- \vec x_j|^2 - \log (1 + g^2 |\vec
x_i- \vec x_j|^2) \right]}\,.
\ee
The equation of motion for a radially symmetric eigenvalue
distribution is therefore ($p \geq 3$)
\be\label{eq:pint}
x = (p-1) \text{Vol} S^{p-2} \int_0^L dr \rho(r) r^{p-1}
\int_0^\pi d\theta \frac{\sin^{p-2}\theta V'(\sqrt{x^2 + r^2 - 2xr \cos\theta}) (x - r \cos
\theta)}{\sqrt{x^2 + r^2 - 2xr \cos\theta}} \,.
\ee
In this expression the potential $V$ is precisely as in
(\ref{eq:potential}) above. As in section \ref{sec:emergence}
above, we can analytically perform the angular integral in
(\ref{eq:pint}) for any given $p$.

\begin{figure}[h]
\begin{center}
\epsfig{file=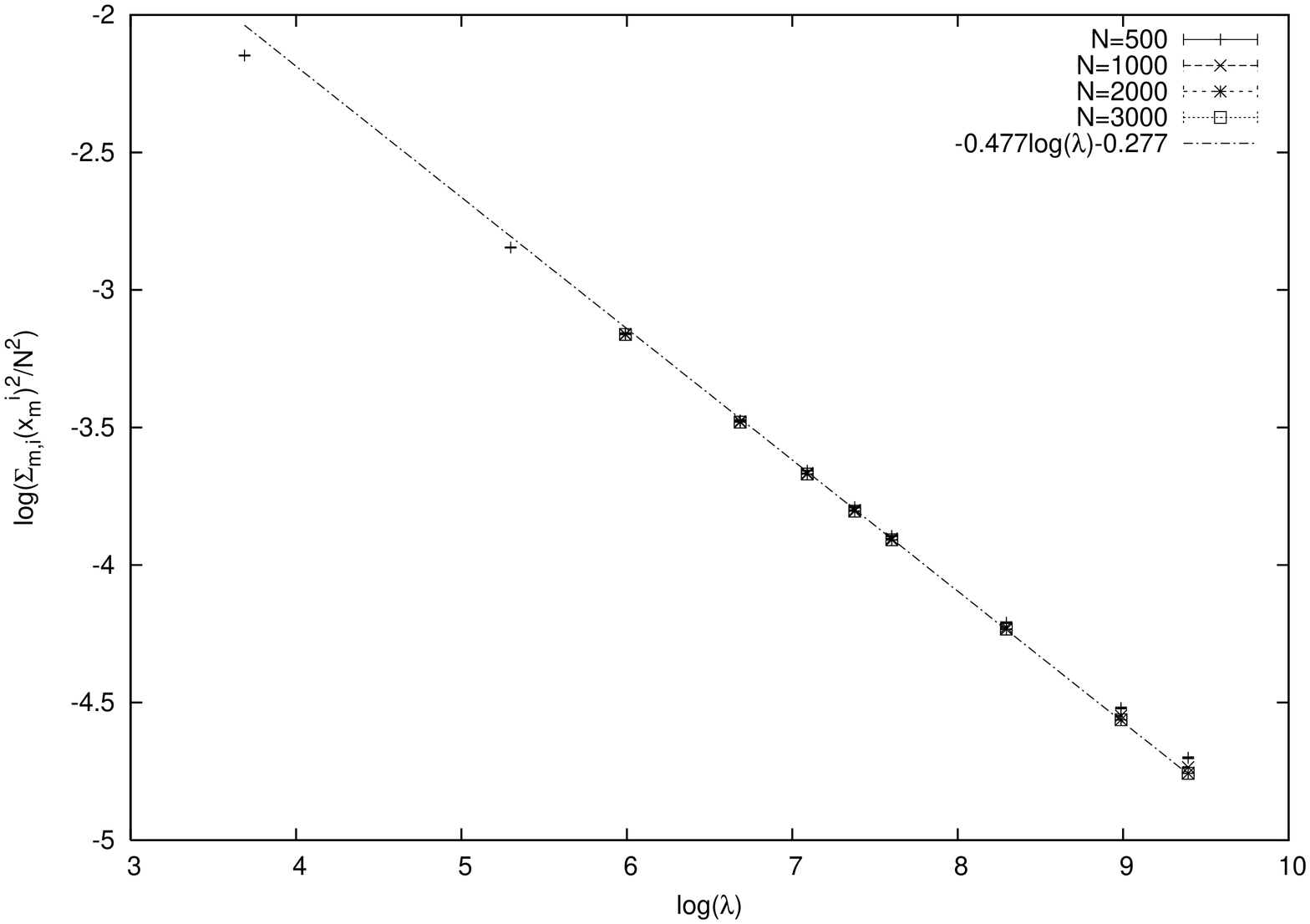,width=2.1in,angle=270,trim=0 0 0 0}%
\epsfig{file=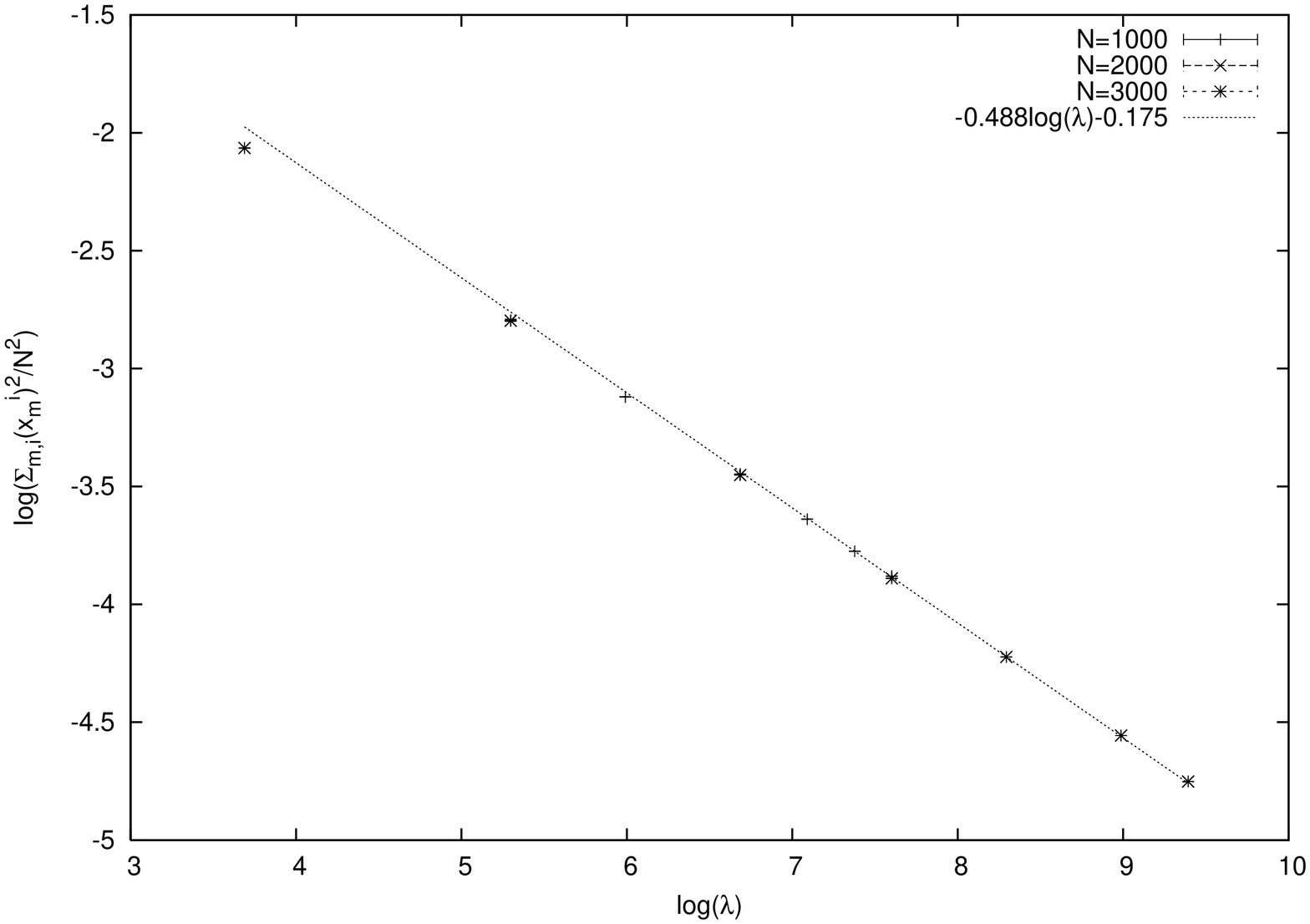,width=2.1in,angle=270,trim=0 0 0 0}%
\end{center}
\caption{For the $p=3$ model (left) an the $p=4$ model (right), plots of $\log(\lambda)$ against
$\log(\sum_{m}\sum_{i=1}^N (x_m^i)^2/N^2)$. The fitting line is
obtained from last two points, $\lambda=8000$ and $\lambda=12000$,
with $N=3000$. \label{fig:3d4d}}
\end{figure}

We have not found an analytic solution for $\rho(r)$ in
(\ref{eq:pint}). It is straightforward to see that if one inserts
a generic ansatz for $\rho(r)$ into the integral equation then the
width of the corresponding distribution will be $L \sim
1/\lambda^{1/4}$. This is on the borderline for selfconsistency of
the one-loop action. Furthermore, we have simulated the commuting
matrix model (\ref{eq:susymass}) to rather high accuracy in $N$
and $\lambda$. The results for the width of the eigenvalue
distribution are shown in figure \ref{fig:3d4d} for the cases of
most interest, $p=3$ and $p=4$.

The results of figure \ref{fig:3d4d} show that the width of the
eigenvalue distribution at strong coupling is a little greater
than $L \sim 1/\lambda^{1/4}$. This is reminiscent of what we
found previously for the commuting bosonic two matrix model in
section \ref{sec:emergence}, where the width was enhanced by a
$\log \lambda$ factor. One possibility is therefore that these
models are indeed commuting and that the eigenvalue dynamics is
not captured by the one loop action (\ref{eq:susymass}). It would
be interesting to test this possibility by simulating the full
partition function (\ref{eq:Zsusymass}) numerically and computing
our observables (\ref{eq:one}) and (\ref{eq:two}). We cannot say
much beyond this for this model with the results we have here.

A curious observation we can make is that if we combine the two
plots of figure \ref{fig:3d4d} into one plot, then the two lines
essentially lie on top of each other.

One new feature that arises in this last model is that there are
fermion zero modes about the commuting saddle. We did not add a
mass term for the fermions and therefore fermionic matrices that
simultaneously commute with the bosonic matrices have zero action.
Therefore the partition function will vanish unless we insert some
fermions to mop up the zero modes. Specifically we should consider
an insertion like
\be
\langle \prod_\a \Psi^\a_{1 1} \cdots \Psi^\a_{N N} \rangle \,.
\ee
We have not written this observable in an explicitly basis
independent way, but rather in the basis given by the commuting
$X$s. The $\Psi$ matrices are anticommuting, so the observable has
to be antisymmetrisable in the components of $\Psi$, as well as
having $N$ entries for each matrix.

\section{Classical fuzzy spheres}

There is another notion of geometry that appears repeatedly in studies of matrix theory and string theory. These are non-commutative spaces formed by the classical matrix degrees of freedom of collections of D branes. For instance, the D0 brane matrices can form a regularized version of a higher dimensional membrane geometry (this was a crucial insight to develop matrix theory \cite{Banks:1996vh}).
Such blowing up into higher dimensions often occurs due to the Myers effect \cite{Myers:1999ps}, and it can also show up in the classification of vacua of some supersymmetric field theories \cite{PS}.

A typical example of such spaces are fuzzy spheres, whose classical coordinates satisfy the $SU(2)$ Lie algebra relations. There are many other such configurations that have been studied. In this section we will make a small study of fuzzy spheres, to complement our points of view from the rest of the paper.

The first thing we would like is a matrix model whose classical saddle points give rise to fuzzy spheres. Define the following matrices, for $i, j, k$ taking values $1,2,3$:
\begin{equation}
M_k = i\epsilon_{ijk} X_i X_j - m X_k \,,
\end{equation}
where the $X$ are hermitian matrices. Now consider the action
\begin{equation}
S= \tr \vec M^2 \,.
\end{equation}
This action can be promoted to a potential for a subsector in a supersymmetric matrix quantum mechanical system (see for example the eleven dimensional plane wave matrix model \cite{BMN}).
A different approach to obtaining an emergent fuzzy sphere from a matrix model may be
found, for instance, in \cite{Steinacker:2003sd, DelgadilloBlando:2007vx}.

If we set $m=0$, classically we have an infinite family of saddles forming a continuous space of solutions. These are characterized by $[X_i,X_j]=0$. This can be interpreted as a classical infrared divergence: the set of classical vacua is a non-compact manifold. There will be zero modes in this system and in principle the classical dynamics can run away along these directions
if one adds some time dependence.

Turning on the parameter $m$ is adding a mass term, plus some other effect linear in $m$. These effects lift the flat directions and one ends up with a finite set of discrete classical vacua.
These vacua are characterized by matrices of order $m$. The general solution is described by $M_k=0$. This can be solved by 
\begin{equation}
X_1 = m  L_1, X_2= m L_2, X_3 = m L_3 \,,
\end{equation}
for some angular momentum representation $\{ L_i \}$ of the Lie algebra of $SU(2)$. Let us choose the $N$ dimensional irreducible representation (the spin $s= (N-1)/2$ representation).

We find in this case that the eigenvalues of $X$ are of order $N$. We can evaluate our criterion for commutativity as a function of $N$.  This is, let us compute the ratios
\begin{equation}
r_1= \frac{- N\tr[X_1,X_2]^2}{\tr X_1^2 \tr X_2^2}, \quad 
r_2= \frac{ \tr (X_1 X_2 X_1 X_2)}{\tr (X_1^2 X_2^2)} \,. 
\end{equation}
The first of these ratios is easy to compute. We get that $i [X_1,X_2] = mX_3$, and using the symmetry between $X_1, X_2, X_3$, we conclude that
\begin{equation}
r_1 = \frac N {\tr L_1^2}=\frac {3}{s(s+1)} \,.
\end{equation}
When we take $N$ to be large this expression vanishes and hence one can argue that the matrices approximately commute.

The ratio $r_2$ is slightly harder to evaluate. We use several identities. Firstly:
\begin{equation}
\tr [L_1,L_2]^2= - N \frac{s (s+1)}3 = 2\tr(L_1L_2L_1L_2) - 2\tr(L_1^2 L_2^2) \,.
\end{equation}
We also find by using symmetries between the $X$ matrices that
\begin{equation}
N\left[ s (s+1)\right]^2=\tr L_2^2= 3 \tr L_1^4+6 \tr (L_1^2 L_2^2) \,,
\end{equation}
and it is easy to show that
\begin{equation}
\tr L_1^4 = \frac 1{15} s(1+s)(1+2s)(-1+3s+3s^2) \,.
\end{equation}
With these results at hand, we compute that
\begin{equation}
r_2= \frac{s^2+s-2}{s^2+s+1/2} = 1 - \frac 5{2s(s+1)+1} \,.
\end{equation}
We see that $r_2$ approaches one as we take $N$ large, again suggesting this can be considered as an approximately commuting matrix model.

In contrast to our previous examples, here the spread of the eigenvalues is of order $N$ rather than $\sqrt N$. This large size can be considered as a non-perturbative effect, because there are many saddles of the matrix model. The claim is that in the large $N$ limit these configurations of large representations go to a smooth geometry, while the vacuum where $X=0$ is non-geometric (as we have seen before).

At this stage, one would also like to understand how to reconcile our picture of large off-diagonal masses causing commutativity with the fuzzy sphere case. Since the gauge group of the configurations is completely broken, and not $U(1)^N$ as in the case of strict eigenvalue saddles, something else must replace this notion. The answer can be found in the work of Bigatti and Susskind \cite{BS}. The idea is that in the presence of some background non-commutativity, the modes that have large momentum become extended, with an extension proportional to the momentum and the noncommutativity. It is clear that here the momentum should be replaced by the angular momentum of fluctuations. We should therefore find that the fuzzy spherical harmonics correspond to stretched segments whose size grows with the angular momentum.  The mass
of these fluctuations also ends up being proportional to this angular momentum. This spectrum has been computed for various types of fuzzy sphere configurations (see for example \cite{BerCorr,Ishiki:2006yr, Kaneko:2007ui}).  It is clear that the modes with very high angular momentum should be very extended and massive, and for the most part the picture will not look too different from the massive off-diagonal modes that we have discussing so far in the other quantum models that we have solved. This seems to suggest that within string theory and matrix models, different notions of emergent geometry may be a lot closer than would appear at first sight.

\section{Summary and discussion}

The basic questions underlying this paper are as follows: suppose
we are given a multi-matrix model that we cannot solve exactly.
Two questions we might ask are the following. Do the degrees of
freedom describing the model reduce in a strong coupling expansion
from order $N^2$ to order $N$? Are these eigenvalue excitations
governed by a simple (e.g. one loop) effective action?

We have found that the answer to these questions depends on the
details of the model. In the cases we could solve exactly, we
found that the observables
\be\label{eq:observables}
\frac{N \tr [X,Y]^2}{\tr X^2 \tr Y^2} \qquad \text{and}
\qquad \frac{\tr XYXY}{\tr X^2
Y^2}\,,
\ee
captured the property of whether the model became commuting at
strong coupling. The first of these observables tends to zero for
all pairs of matrices in a commuting model whereas the second
tends to one. We found that this occurred in a bosonic two matrix
model with a commutator squared interaction. We showed that this
model is described at strong coupling by an emergent two
dimensional hemisphere geometry. The eigenvalue dynamics about
this geometry, however, is not describable by the one loop
effective action.

For the other solvable models we considered there was no emergent
geometry at strong coupling. In the bosonic model of section
\ref{sec:boson} this occurred because integrating out the off
diagonal modes induced a strong attractive force on the
eigenvalues. The clumped together eigenvalues then had light
off-diagonal modes connecting them. In the fermionic model of
section \ref{sec:fermion} this attractive force was cancelled, but
the lack of interactions between some of the matrices (necessary
to solve the model exactly) meant that these matrices were
uncorrelated and did not commute.

We then turned to fully interacting models which we cannot solve
exactly. For the bosonic model we argued that there was no local
emergent geometry. For models with fermions we found apparently
consistent commuting saddles in a model in which the fermionic
measure was constructed by hand to cancel the bosonic one loop
effective action. These models appear to have an emergent
spherical geometry. It would be very interesting to test our
(self-consistent) results for that model with a full numerical
simulation of the partition function (\ref{eq:fullfermions}). For
the supersymmetric model with a bosonic mass term
(\ref{eq:Zsusymass}) we did not find clearcut evidence for
emergent geometry. It would also be very interesting to simulate
that model numerically and to compute our observables
(\ref{eq:observables}) at strong coupling.

We also considered some classical fuzzy spaces as other toy models where geometry 
can be argued to appear. We found that in these cases, at the classical level one also found that the 
matrices making up the fuzzy sphere are close to being commuting in the large $N$ limit.
The size of the matrices ends up being of order $N$ rather than $\sqrt N$, and this is what makes them more classical. We also found that in these systems the notion of off-diagonal modes should be captured by the spherical harmonics of the fuzzy sphere. The 
angular momentum should be correlated with the size of the segment joining two 
putative points on the sphere.

An important feature of all the multi-matrix models (both quantum and classical) 
that we have
considered in this paper is that they have a mass term. This is
very natural in AdS/CFT, because the boundary of $AdS$ in global
coordinates contains a spatial three-sphere which gives even the
zero modes a conformal mass. It is different, however, from
previous considerations of matrix models as putative
nonperturbative formulations of string theory, as these do not usually
have mass terms. In the examples we considered, a balance between
the mass term and the commutator squared interaction term was
important for the emergence of local commuting geometry at strong
coupling. Otherwise the coupling dependence is simply dimensional
analysis. In particular, the cases with emergent geometry occur
when the massless model has infrared divergences. These infrared divergences signal a potential instability of the system to grow in size. It is the regulation of this growth that seems to give us a notion of geometry. If the size is not controlled, one can imagine that the end of such a scenario is a system where all the eigenvalues have scattered at infinite distance from each other and there is nothing interesting left. We would definitely not want to call such type of configuration a geometry, but the dynamical process of reaching such an end configuration could be a cosmology of sorts. Previous
cosmological applications of matrix models include \cite{Freedman:2004xg,Craps:2005wd,Erdmenger:2007xs}.

We believe it is worth exploring these ideas further. There are various supersymmetric matrix quantum mechanical models where there is a mass term (for example, the plane wave matrix model \cite{BMN}), and our notions of geometry might be useful to tackle the spectrum of extended objects in such systems.  
In a different direction, recent progress has been made connecting black hole physics with supersymmetric matrix models at finite temperature (with infrared divergences) \cite{Catterall:2007fp, Anagnostopoulos:2007fw, Catterall:2008yz}. Infrared divergences can also be found in weakly coupled
${\mathcal{N}}=4$ Super Yang-Mills theory on a three sphere, e.g. \cite{Hollowood:2008gp}. At high temperatures one expects off diagonal modes to play an important part in the dynamics. Therefore one faces an interesting challenge in combining our notion of an emergent geometry with black hole physics.

\section*{Acknowledgements}

We would like to thank Ofer Aharony, Jun Nishimura and Daniel
Robles-Llana for discussions. SAH would like to acknowledge the
hospitality of the Weizmann Institute, where this project was
initiated. Simulations were performed using the YITP cluster
system. This research was supported in part by the National
Science Foundation under Grant No. PHY05-51164, and by the DOE
under grant DE-FG02-91ER40618.

\appendix

\section{Some corrections away from large $\lambda$}
\label{sec:corrections}

\subsection{Correction to the parabolic distribution}

In order to compute the four point functions in section
\ref{sec:2matrix} to subleading order, we needed the leading order
correction to the large $\lambda$ eigenvalue distribution
\be\label{eq:total}
\rho(x) = \rho_0(x) + \frac{\rho_1(x)}{\l^{1/3}} \,.
\ee
Here $\rho_0(x)$ is the parabolic distribution
(\ref{eq:solution}). It turns out that for the leading order
correction we can take $L$ to remain given by (\ref{eq:L}) and
$\int \rho_1(x) dx = 0$, thus retaining $\int \rho(x) dx = N$.

Using (\ref{eq:general}) and expanding the equations of motion
(\ref{eq:integral}) to second order in strong coupling, we find
the following expression for the correction to the distribution
\be\label{eq:rho1}
\rho_1(x) = \frac{N}{L^2} \left(\frac{2}{3\pi}\right)^{1/3}
\frac{3}{4\pi} \left(x \log \frac{L-x}{L+x} + L \right) \,.
\ee
Clearly this solution is somewhat formal, as the eigenvalue
density diverges as $x \to \pm L$. However, inside the integral it
does solve the expanded equation of motion. Furthermore, the
region in which the full distribution (\ref{eq:total}) becomes
large is exponential small ($\sim e^{- \l^{1/3}}$). This is
related to the fact that the parabolic solution $\rho_0(x)$ was
not valid very close to the endpoints, as we noted in the main
text. Therefore, we should be able to use (\ref{eq:rho1}) to
evaluate observables for which the integral over the eigenvalues
is finite.

Using the integrals in (\ref{eq:XYXY}) and (\ref{eq:XXYY}) it is
easy to use (\ref{eq:rho1}) to find the correction to the four
point functions that is due to $\rho_1(x)$:
\be
\delta (\tr XYXY) = \delta (\tr X^2 Y^2) = - \frac{N^3}{40 \l} \,.
\ee

\end{document}